\documentclass[sigconf]{acmart}

\usepackage{algorithm,algorithmic}
\usepackage{balance}

\AtBeginDocument{%
  }

\copyrightyear{2024}
\acmYear{2024}
\setcopyright{acmlicensed}\acmConference[CIKM '24]{Proceedings of the 33rd ACM International Conference on Information and Knowledge Management}{October 21--25, 2024}{Boise, ID, USA}
\acmBooktitle{Proceedings of the 33rd ACM International Conference on Information and Knowledge Management (CIKM '24), October 21--25, 2024, Boise, ID, USA}
\acmDOI{10.1145/3627673.3680054}
\acmISBN{979-8-4007-0436-9/24/10}
\begin{document}

\title{A Real-Time Adaptive Multi-Stream GPU System for Online Approximate Nearest Neighborhood Search}

\author{Yiping Sun}
\authornote{Corresponding Author.}
\email{sunyiping@xiaohongshu.com}
\orcid{0009-0000-9559-8268}
\affiliation{%
  \institution{Xiaohongshu Inc}
  \city{Shanghai}
  \country{China}
}

\author{Yang Shi}
\email{shiyang1@xiaohongshu.com}
\orcid{0009-0009-2042-5405}
\affiliation{%
  \institution{Xiaohongshu Inc}
  \city{Shanghai}
  \country{China}}

\author{Jiaolong Du}
\email{jiaolong@xiaohongshu.com}
\orcid{0009-0006-7993-4755}
\affiliation{%
  \institution{Xiaohongshu Inc}
  \city{Shanghai}
  \country{China}
}

\renewcommand{\shortauthors}{Yiping Sun, Yang Shi, \& Jiaolong Du}

\begin{abstract}
In recent years, Approximate Nearest Neighbor Search (ANNS) has played a pivotal role in modern search and recommendation systems, especially in emerging LLM applications like Retrieval-Augmented Generation. There is a growing exploration into harnessing the parallel computing capabilities of GPUs to meet the substantial demands of ANNS. However, existing systems primarily focus on offline scenarios, overlooking the distinct requirements of online applications that necessitate real-time insertion of new vectors. This limitation renders such systems inefficient for real-world scenarios. Moreover, previous architectures struggled to effectively support real-time insertion due to their reliance on serial execution streams.
In this paper, we introduce a novel Real-Time Adaptive Multi-Stream GPU ANNS System (RTAMS-GANNS). Our architecture achieves its objectives through three key advancements: 1) We initially examined the real-time insertion mechanisms in existing GPU ANNS systems and discovered their reliance on repetitive copying and memory allocation, which significantly hinders real-time effectiveness on GPUs. As a solution, we introduce a dynamic vector insertion algorithm based on memory blocks, which includes in-place rearrangement. 2) To enable real-time vector insertion in parallel, we introduce a multi-stream parallel execution mode, which differs from existing systems that operate serially within a single stream. Our system utilizes a dynamic resource pool, allowing multiple streams to execute concurrently without additional execution blocking. 3) Through extensive experiments and comparisons, our approach effectively handles varying QPS levels across different datasets, reducing latency by up to $40\%$-$80\%$. The proposed system has also been deployed in real-world industrial search and recommendation systems, serving hundreds of millions of users daily, and has achieved significant results.
\end{abstract}




\maketitle

\section{Introduction}
The task of Approximate Nearest Neighbor Search (ANNS) \cite{abbasifard2014survey} has gained extensive application and research in modern contexts, such as search systems, recommendation systems, and information enhancement for Large Language Models (LLM), especially Retrieval-augmented generation (RAG) \cite{gao2023retrieval}. Its objective is to efficiently identify approximate Top-K results through approximate algorithms like clustering \cite{nolet2023cuslink}, quantization \cite{jegou2010product}, tree-based \cite{silpa2008optimised}, and graph-based methods \cite{malkov2018efficient}. These methods, while sacrificing a marginal recall rate, significantly reduce the computational costs associated with distance calculations. This enables the expedited retrieval of approximate Top-K results within considerably large datasets in a shorter timeframe.
Furthermore, even with the adoption of efficient approximate algorithms, ANNS systems based on CPUs still face the challenge of high computational load and the bottleneck of modern CPU memory bandwidth. An increasing body of research is turning to the high computational power and memory bandwidth of GPUs to address the inadequacies in CPU performance. In the system Faiss \cite{IEEEBILLION-SCALE} and Raft \cite{rapidsai}, the concept of utilizing GPUs for retrieval instead of CPUs was first introduced, achieving an huge performance improvement compared to CPUs. The implementation details are openly available and have gained widespread usage in the industry. 

However, we have identified that the aforementioned systems are only designed for offline scenarios, where vectors in the index are processed offline and loaded before performing search. These systems are unable to address the most common online scenarios in Approximate Nearest Neighbor Search (ANNS), which require high-frequency real-time insertions of vectors. This capability is crucial for  search/recommendation applications \cite{gong2022real}, where immediate updates ensure the most recent data is always available for accurate and relevant search results. These limitations arise from two main aspects. 

\begin{figure*}[htbp]
\centering
\includegraphics[width=0.9\linewidth]{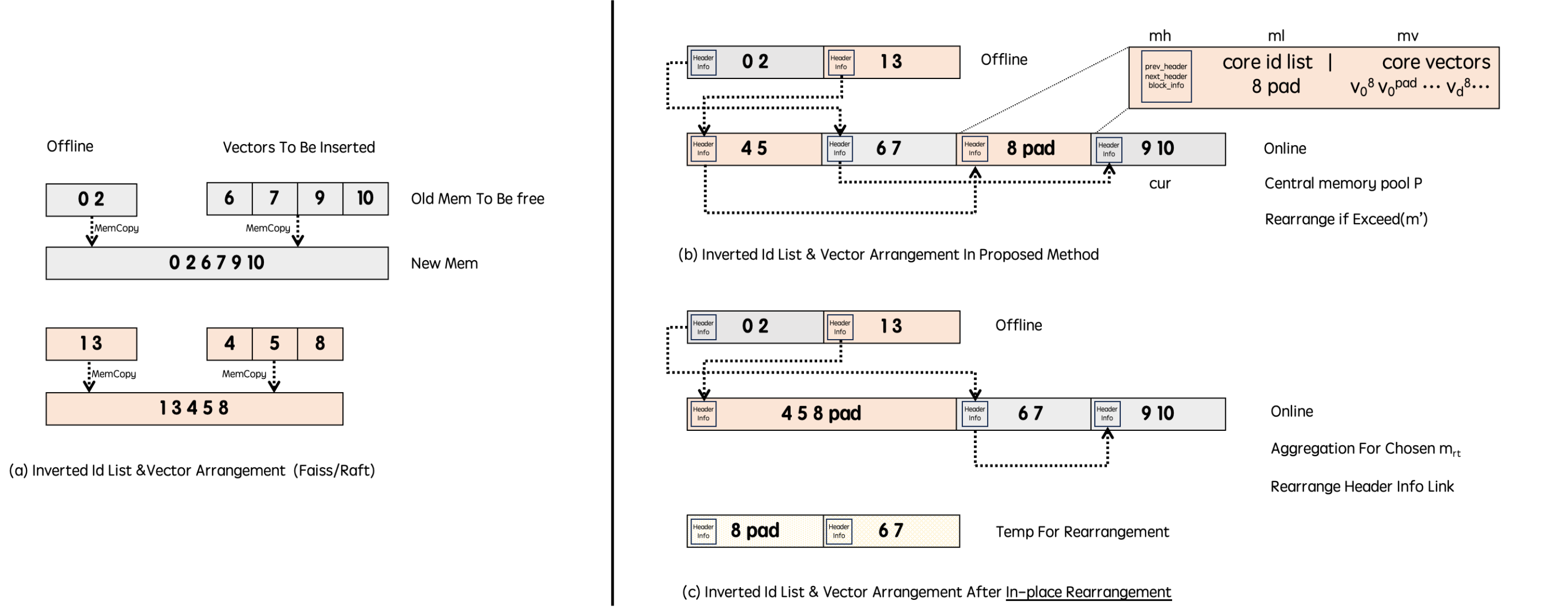}
\caption{(a) Inverted Id List and Vector Arrangement in Faiss/Raft: new vectors are appended, and new memory space is allocated. Old memory space is freed after the new one is ready. (b) Inverted Id List and Vector Arrangement in Proposed Method: the memory block is applied when new vectors need to be inserted. Each memory block has a header indicating its previous and next blocks. In this example, the IDs 0, 2, 6, 7, 9, and 10 can be connected as a single ID list. (c) Inverted Id List and Vector Arrangement after In-place Rearrangement: if the memory block list Exceed($m'$), dynamic rearrangement of fragmented memory blocks are executed. Temporary segments are utilized in this process. In this example, the IDs 4, 5, 8, and "pad" are aggregated, optimizing the header jump from twice to one.}
\label{fig:overview}
\end{figure*}

On the one hand, the memory arrangement of vector insertion is not well designed in existing systems (Faiss/Raft), which almost needs to be merged with existing vectors. This process is illustrated in Figure \ref{fig:overview}a and detailed in Algorithm \ref{alg:extend}. For typical inverted-index-based ANNS algorithms like Ivfflat and Ivfpq, vectors are initially partitioned into multiple clusters (line 3). During the search process, the vector lists belonging to the most relevant clusters to the query are computed, and the final top-k vectors are retrieved. When new vectors arrive, existing systems typically needs to copy the vector lists of the new vectors from the offline segment to the new segment (line 5 - 14). Such method, therefore, incurs significant overhead on the GPU, as both memory allocation and copying require triggering a kernel launch which consumes substantial resources. Additionally, it should be noted that in this article, we only discuss inverted-based ANNS algorithms, as graph-based ones \cite{wang2021comprehensive} are currently challenging to support in real-time, even on CPUs.

\begin{algorithm}[t]
    \caption{Ivfflat and Ivfpq Approximate Online Vector Operations on Existing GPU Systems (Faiss/Raft)}
    \label{alg:extend}
    \begin{algorithmic}[1]
        \REQUIRE offline vectors $x_i \in X$, new coming online vectors $y_i \in Y$ and number of new vectors $M$
        \REQUIRE $m_k.size$ for new allocated extra memory space size, $wl_k$ for temporary vector list assignment
        \REQUIRE vector lists $l \in L$, corresponding cluster $c \in C$, number of vector lists/clusters $N$, $C$ is generated by Kmeans($X, N$)
        \STATE // Calculate belonged vector list, GPU Block Execution
        \FOR{$i$ from $0$ to $M - 1$}
            \STATE $k \leftarrow \min(c, y_i)$  // where $\forall c \in C$
            \STATE $m_k.size \leftarrow  m_k.size + 1$
            \STATE $\text{append}(wl_k, y_i)$
        \ENDFOR
        \STATE // Append to new lists, GPU Block Execution
        \FOR{$j$ from $0$ to $N - 1$}
            \IF{Reallocation needed}
            \STATE Allocate new space $l_j'$ of $l_j.size + m_k.size$ allocated
            \STATE ${l_j'} \leftarrow Merge(l_j, wl_j)$
            \STATE free $l_j$ and $l_j \leftarrow l_j'$
            \ENDIF
        \ENDFOR
        \RETURN purely new and complete id lists $L$
    \end{algorithmic}
\end{algorithm}

On the other hand, existing systems employ a single-stream serial execution mode for ANNS. In scenarios where tasks are singular and dependent on each other (e.g., the current kernel relies on the output of the previous kernel), the execution is highly ordered, with data processed one by one as a batch to maximize throughput. However, in cases where GPU tasks are diverse and independent, this execution mechanism may not suffice for handling complex scenarios. As illustrated in Figure \ref{fig:multi-stream}, systems that rely on serial execution need to wait separately for real-time vector extension. If the volume of data being copied is substantial, this waiting period can be prolonged, which could block the execution of the incoming search. For example, in the case of a long vector list (hot data), a single real-time vector insertion could take tens of milliseconds due to repeated copying, thereby obstructing normal retrieval operations.

So we want to design such a system to tackle the above shortcomings: 1) online vector insertions can be updated real-time 2) online vector insertions require the least extra memory 3) online vector insertions can be processed parallel with online  search procedure.

In this article, we introduce a novel \textbf{R}eal-\textbf{T}ime \textbf{A}daptive \textbf{M}ulti-\textbf{S}tream System for Online \textbf{G}PU \textbf{A}pproximate \textbf{N}earest \textbf{N}eighbor \textbf{S}earch (\textbf{RTAMS-GANNS}). Our innovation includes a well-designed dynamic vector operation algorithm along with a novel execution mode which leverages the parallelism inherent in GPU streams, enabling a high-performance concurrent execution of vector search and real-time vector insertion. Our contributions can be summarized as follows:

\begin{itemize}
    \item We propose an innovative approach to handle real-time vector insertion, utilizing a dynamic algorithm based on memory blocks. This method not only allows for the seamless integration of new vectors but also incorporates in-place rearrangement, optimizing memory utilization. By minimizing the need for costly GPU memory allocation and copying, our solution ensures efficient real-time insertions without compromising performance.
    \item In addition to the dynamic vector insertion algorithm, we introduce a multi-stream parallel execution mode to enable real-time insertions. Unlike traditional systems that execute tasks sequentially within a single stream, our architecture embraces parallelism by utilizing a stream-cached dynamic resource pool. This innovative design enables different streams to execute concurrently, eliminating the need for additional execution blocking. 
    \item Extensive experiments and comparisons validate the effectiveness of our approach in benchmark dataset and industrial dataset. Our solution  effectively handles varying QPS levels across different datasets, reducing latency by up to $40\% - 80\%$. With successful deployment in real-world industrial search and recommendation systems, our solution has proven instrumental in serving hundreds of millions of users daily, yielding remarkable outcomes.
\end{itemize}
\section{Related Work}
\subsection{GPU facilitated Approximate Nearest Neighbour Search}
As mentioned earlier, Faiss \cite{IEEEBILLION-SCALE} and Raft \cite{rapidsai} are pioneering works that elevated vector retrieval systems from CPU to GPU, inaugurating the era of GPU-accelerated Approximate Nearest Neighbor Search (ANNS). Subsequent endeavors have built upon this foundation, focusing on either algorithmic enhancements or system architecture. In the realm of algorithms, RobustiQ \cite{ICMR19-SIGCONF} optimized and extended quantization algorithms for GPU implementation, while works such as \cite{CAGRA}\cite{yu2022gpu}\cite{SONG} made significant strides in improving GPU-based graph algorithms. Regarding system design optimization, \cite{karthik2024bang}\cite{online1}\cite{zhang2024fast} independently converged on integrating heterogeneous CPU-GPU scheduling, distributing data between GPU and CPU, and establishing periodic synchronization mechanisms. Despite these optimizations, all existing approaches still rely on a serial offline system and fall short of effectively addressing the challenges posed by real-time vector insertions.

In this study, we shift our focus from theoretical algorithmic improvements to system architecture. We present the first GPU ANNS system operating on a parallel architecture, providing an effective solution for online scenarios and marking a clear distinction from the aforementioned systems. This approach addresses the critical need for high-frequency real-time vector insertions, ensuring up-to-date and accurate search results in dynamic environments.

\subsection{Online Vector Insertion Methods in CPU/GPU ANNS}
In CPU systems, inserting a vector into memory is a straightforward process. For inverted-based algorithms, SPFresh \cite{xu2023spfresh} proposed an incremental rebalancing tool to split vector partitions and reassign vectors in nearby partitions, adapting to shifts in data distribution. Since this algorithm is applied in CPU systems, their optimization focus differs from ours. They mainly focus on optimizing the balance of the id list, which is another real-time concern while we are concerned with reducing the expensive memory allocation overhead, non-aligned memory access, and blocking execution frameworks in GPUs. Therefore it is still difficult to apply this methods on GPUs without a feasible framework before our system is proposed. For graph-based algorithms, FreshDiskANN \cite{aditi2021freshdiskann} designed a novel two-pass Streaming-Merge algorithm for efficiently merging the in-memory index with the SSD-index. This algorithm is designed for SSD-based disk indexing, proposing real-time update methods primarily focused on updates under large data scenarios rather than making the update mechanism more tailored to disks. Additionally, due to significant differences between GPU and CPU graph algorithms, this method cannot be directly applied to GPUs. In this paper, we only focus on the currently more widely-used IVF methods instead of graph based methods. In addition, \cite{realtime1}\cite{karthik2024bang} focus on using heterogeneous hardware, combining CPUs and GPUs to facilitate computations. While they may load the latest vectors during the process of exchanging data from CPU to the GPU, these methods themselves do not aim to address the real-time vector insertion problem. Moreover, they still encounter issues with expensive copying and reallocation, as illustrated in the previous section, making such approaches difficult to implement in practice. 

\section{System Overview}
\subsection{Memory Block Based Real-time Vector Insertion Algorithm}
In this section, we will primarily introduce our proposed memory block based dynamic vector insertion algorithm. The overview memory layout is depicted in Figure \ref{fig:overview}b and the main algorithm process is illustrated in Algorithm \ref{alg:core}. Instead of arranging all vectors in continuous memory as in existing systems, we link them as a list using the header of each memory block $m$ for the vector id list $l$ of a certain cluster $c$ in ivfflat and ivfpq, which are represented by the same color in Figure \ref{fig:overview}b. To mitigate memory allocation overhead \cite{winter2021dynamic} during processing, we have designed a memory allocation method in Algorithm \ref{alg:core} (line 13). It comprises the central memory pool $P$ (occupying almost the entire GPU memory) and the memory block $m$ (the smallest unit). The central memory pool $P$ has been pre-split by memory blocks, which are labeled from 0, 1 ... to $|P|$. The $cur_P$ represents the current position of latest allocated memory block. When allocating a memory block, we would atomicAdd the $cur_P$ and locate the memory block logically without any real extra memory allocation.
\begin{algorithm}[t]
    \caption{Memory Block Based Dynamic Vector \textbf{Insertion}}
    \label{alg:core}
    \begin{algorithmic}[1]
        \REQUIRE offline vectors $x_i \in X$, new coming online vectors $y_i \in Y$
        \REQUIRE vector id lists num $nl \in NL$, corresponding cluster $c \in C$, number of clusters $N$, $C$ is generated by Kmeans($X, N$)
        \REQUIRE memory block $m \in M$, memory block list $m' \in M'$
        \REQUIRE central memory pool $P$, latest allocated memory block position $cur_P$
        \STATE $gid = \sum_{nl}nl , \forall nl \in NL$
        \STATE // Insertion
        \STATE // GPU Multi Thread Execution, parallel by $y_i$
        \FOR{$i$ from $0$ to $|Y| - 1$}
            \STATE $k \leftarrow \min(c, y_i)$  , $\forall c \in C$
            \STATE $did = atomicAdd(nl_k, 1)$
            \STATE $mid = did \ / \ T_m$, $mo\!f\!f = did \ \% \ T_m$
            \STATE $m_{mid} \leftarrow$ last memory block in $m_k'$
            \STATE \_syncthreads() // make sure $m_k'$ is updated later
            \IF{$mid >= |m_k'|$}
                \STATE inNewBlock $\leftarrow$ true
                \IF{$mo\!f\!f == 0$}
                \STATE $m_{mid} \leftarrow P[atomicAdd(cur_P, 1)]$
                \STATE link $m_{mid}$ with last block in $m_k'$
                \ENDIF
            \ENDIF
            \STATE \_syncthreads() // make sure new $m_k'$ can be accessed 
            \IF{inNewBlock}
                \STATE $m_{mid} \leftarrow$ new last memory block in $m_k'$
            \ENDIF
            \STATE fill vector in $m_{mid}$ by dimension 32 interleave layout
        \ENDFOR
        \STATE // Rearrange
        \STATE // GPU Single Thread Execution
        \FOR{$i$ from $0$ to $|M'| - 1$}
            \IF{Exceed($m_i'$)}
                \STATE \textbf{Rearrange($m'$)}
            \ENDIF
        \ENDFOR
        \RETURN updated memory block $m$
        \end{algorithmic}
\end{algorithm}
The entire algorithm relies on the smallest unit memory block $m$. A vector id list $l$ of memory blocks for newly inserted vectors is defined as $m'$:
\begin{equation}
m' = [m_{i_0}, m_{i_1}, ..., m_{i_n}] \quad \forall i \in l
\end{equation}
For each memory block $m$, it may contain $|ml|$ vectors, which of limit $T_m$ can be pre-set by users. The structure of a memory block is also illustrated in Figure \ref{fig:overview}b. It consists of a header $mh$, core id list $ml$, and core vectors $mv$. The header $mh$ contains important information that indicates the address of the prev\_header, next\_header and block\_info (vector capacity, vector size,  etc.). The core id list $ml$ saves the vector ids belong to the memory block of this cluster. The core vectors $mv$ in each memory block are arranged interleaved by 32 dimensions, similar to Faiss/Raft, determined by the thread number in a GPU warp to enable coalesced read instructions.
\begin{equation}
mv = [mv_0^0, mv_0^1, ..., mv_0^{31}, mv_1^0, ..., mv_d^i, ...] \ \forall d \in D, i \in ml
\end{equation}
where $D$ is the dimension of the single vector. When a new vector arrives, the algorithm will first call \textbf{Insertion} to dynamically insert the vector into the memory block. When executing insertion, the algorithm is executed parallel on new coming online vectors to get the maximum parallel performance. We first calculate the belonged cluster of newly inserted vector followed by ivfflat and ivfpq (line 5) and then calculate whether this vector should be added to a new linked memory block or an existing memory block represented by did (line 6). When vectors need to be added to a new memory block because the existing block is full, conflicts can occur in a multi-threaded environment, potentially causing issues with concurrent access. So we designed a lock-free allocation method \cite{5577907}. We first find the location of the newly inserted vector, which is determined by memory block id $mid$ and offset in this memory block $mo\!f\!f$ by using an atomic value indicating the length of vector lists $nl$ (line 6 - 8). Then, for threads that need to allocate a new memory block, we call allocation of $m_{mid}$ (line 13),  relink the memory block list $m_k'$ and fill the vectors (line 10 - 20). As stated before, the allocation is also thread-safe which just calculates a logic location of the memory block. After insertion, if the length of a certain newly inserted memory block list $m'$ exceeds a predefined value $T_{m}'$, \textbf{Rearrangement} should be executed (line 23 - 27).
\begin{equation}
    \text{Exceed}(m') = \sum_{m}|m.ml| > T_{m}' \quad \forall m \in m'
\end{equation}
The rearrangement algorithm is illustrated in algorithm.\ref{alg:rearrange}. The core idea of rearrangement is to merge two split memory block together so that the vectors are continuous and a header jump is eliminated. In a certain memory block list $m'$, we exchange the memory block next to the first memory block with linked memory block using a extra temp memory segment and then update the first block header. It may happen that the memory block we want to exchange is a merged memory block, under certain circumstance, we should first split it and do merge on it recursively or iteratively.

\begin{algorithm}[htbp]
    \caption{Memory block \textbf{Rearrangement}}
    \label{alg:rearrange}
    \begin{algorithmic}[1]
        \REQUIRE memory block $m \in M$, memory block list $m' \in M'$
        \STATE $m_i \leftarrow$ memory block of $m'$ to be extended
        \STATE $m_{j} \leftarrow$ memory block of $m'$ to extend
        \IF{$m_{i+1}$ is merged}
            \STATE split($m_{i+1}$)
            \STATE lazyMerge $\leftarrow$ true
        \ENDIF
        \STATE prepare tmp segment for $m_{i+1}$ and $m_{j}$
        \STATE wait for $m_{i+1}$ and $m_{j}$ is spare
        \STATE $m_{i+1}$.copyFrom($m_{j}$)
        \STATE $m_j$.copyFrom(tmp $m_{i+1}$)
        \STATE update header of $m_{i+1}$ and $m_j$
        \STATE wait for tmp $m_{i+1}$ and $m_{j}$ is spare
        \IF{lazyMerge}
            \STATE \textbf{Rearrange}(new $m_j$) // old $m_{i+1}$
        \ENDIF
    \end{algorithmic}
\end{algorithm}

\subsection{Stream-Cached Multi-Stream Execution}

\begin{figure}[t]
  \centering
  \includegraphics[width=\linewidth]{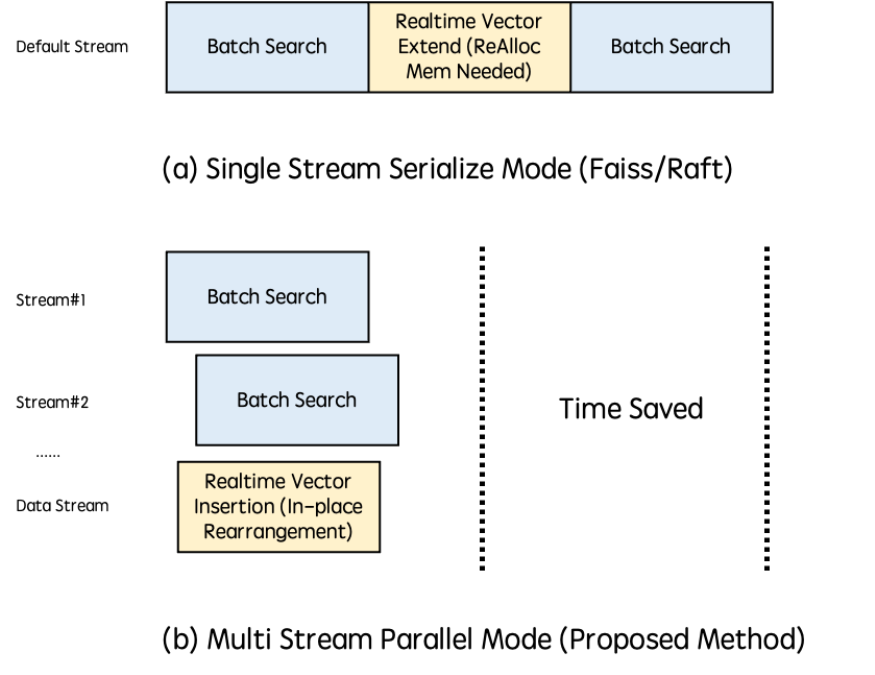}
  \caption{(a) Single Stream Serialize Mode (Faiss/Raft): execute kernel in a single-stream, cannot handle scenario including new coming vectors. (b) Multi Stream Parallel Model (Proposed Method): enabling kernel execution under a parallel form, not only improving online gpu utilization but also adapting real-time vector operations naturally.}
  \label{fig:multi-stream}
\end{figure}

In this section, we propose a parallel execution mechanism to enable real-time vector insertion kernels to run concurrently with normal retrieval sorting kernels on the GPU. As depicted in Figure \ref{fig:multi-stream}a, previous systems operated in a serialized manner—a common practice in GPU applications due to the massive computational workload, which maximizes throughput by running a single kernel at a time \cite{pinnecke2015toward}. However, in ANNs, individual request computations may not fully utilize the GPU's performance, and the serial execution of real-time vector insertion kernels can block normal retrieval sorting kernels. Therefore, a parallel system is more suitable for online systems. By concurrently executing these kernels, we not only reduce significant execution time (Time Saved in Figure \ref{fig:multi-stream}(b)) but also make real-time vector insertion more immediate.
We adopt multi-stream parallelism \cite{czarnul2020investigation} in CUDA, which needs domain knowledge to design a kernel execution mechanism. All kernels submit to gpu should be run on a certain stream. As shown in algorithm.\ref{alg:multi-stream}, for Search kernel, each kernel will run on a distinct stream and the stream resource can be reused through a resource pool. For Insertion kernel, it should be run on a same data stream. So finally it will have multiple streams for Search and single stream for Insertion. However, for successful execution, the system needs to be considered more: 1) The system must avoid online memory allocation and deallocation to prevent global blocking and degradation into a serialized mechanism. 2) Each kernel in the system should not consume a significant amount of resources, as excessive resource consumption would lead to block.

For consideration 1, drawing inspiration from methods employed in TCMalloc \cite{manghwani2011scalable}, we devised a dual-layered stream-based resource pool. Initially, each request for batch search is allocated to a dedicated stream, possessing a separate small memory allocation (sufficient for the use of ivfflat and ivfpq algorithms). Additionally, when the demand for resources is substantial, a larger resource allocation will be requested from the central memory pool, which is then returned after use. For the real-time vector insertion kernel, we considered parallelism and lock-free allocation of GPU memory during the design phase, as outlined in Algorithm \ref{alg:core}(line 13). Consequently, it will not encounter any blocking issues and can smoothly parallelize with the retrieval kernel.

For consideration 2, since multiple kernels are scheduled on stream multiprocessors (SM) simultaneously \cite{wang2016simultaneous}, it's essential that a single kernel doesn't monopolize all computational resources \cite{zhong2013kernelet}. Additionally, when the number of threads is large, there will be limitations on the number of threads per block on the GPU. Therefore, we will restrict the number of blocks and threads per kernel, which may result in increased latency for online requests with high computational demands. However, this is quite common in online scenarios.
\begin{algorithm}[htbp]
    \caption{Multi Stream Execution}
    \label{alg:multi-stream}
    \begin{algorithmic}[1]
        \REQUIRE Resource Pool $R$, Central Memory Pool $P$
        \REQUIRE Request $q$
        \IF{q.type == batch search}
            \STATE $r$ (stream, memory for stream) $\leftarrow$ Resource Pool $R$
            \IF{memory is not enough}
                \STATE $r$.memory $\leftarrow$ Central Memory Pool $P$ 
            \ENDIF
            \STATE submit kernel Search(q.queries, $r$) on $r$.stream
        \ELSIF{q.type == vector insertion}
            \STATE submit kernel Insertion(q.vectors) on data stream
        \ENDIF
    \end{algorithmic}
\end{algorithm}

\subsection{Deployment Detail}
In this section, we delve into the deployment intricacies of our real-time adaptive multi-stream GPU system within real-world systems. Our system has been operational for over six months on T4/A10 GPU machines, seamlessly integrated into both the search and recommendation systems of a widely-used information app catering to over 100 million daily users. Comprising two essential segments, namely the offline and online components, our system efficiently manages both prepared data and real-time vector insertion tasks.

The offline segment is dedicated to processing prepared data, while the online segment orchestrates real-time vector insertion operations. Managed by multiple CPU threads and interfacing with Kafka for message ingestion, the online segment implements a dynamic batching strategy, aggregating vectors either every second or upon reaching a threshold at multiples of 128 insertions, with a cap at 1024, before dispatching the aggregated batch to the GPU.

Before kernel execution, we meticulously allocate 32 independent resources from the resource pool, each provisioned with 50MB of cached memory, as previously detailed. In instances where temporary memory requirements surpass this threshold, additional memory is sourced from the central pool, with each allocation set at 200MB. To ensure equitable resource allocation, particularly under high QPS conditions, we have devised a lock-free queue mechanism. Requests are rejected when all 32 resources are exhausted. Additionally, we establish a dedicated stream for vector insertion tasks to streamline processing.

For the central memory pool \( P \), we employ it across two distinct domains: search and insertion. To maintain cohesive control within a singular pool, we adopt a high address allocation strategy for search and low addresses for insertion. An alert mechanism triggers when resource utilization exceeds the $90\%$ threshold.
\begin{figure*}[t]
\centering
\captionsetup{skip=5pt}
\includegraphics[width=0.86\linewidth]{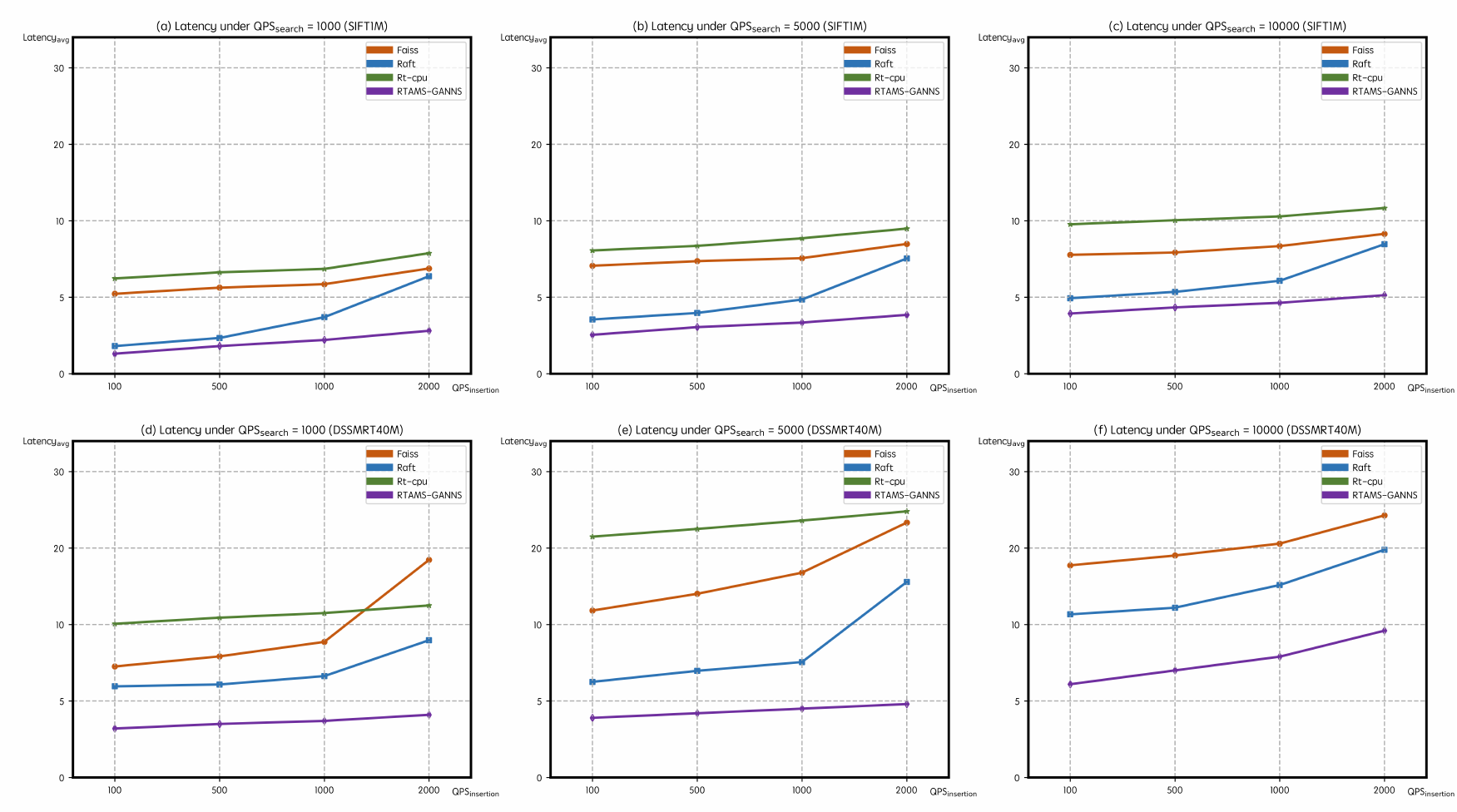}
\caption{Latency Comparison on SIFT1m and DSSMRT40M under QPS$_{search}$ = 1000, 5000, 10000}
\label{fig:compare}
\end{figure*}
During kernel execution, memory allocation for each block \( m \) is meticulously tailored to accommodate 1024 vectors, with thread counts deliberately set below 1024 to preempt conflicts. This proactive measure mitigates scenarios where simultaneous requests for two new memory blocks for vector insertion arise. At present, in both the vector insertion and rearrangement kernels, only one block is activated to circumvent potential resource contention inherent in the current design. Nonetheless, this level of parallelism suffices to manage the workload segment effectively.
\section{Experiment}

In this section, we aim to validate the performance of our system in various environments through experimentation. We will conduct tests on both the standard dataset SIFT1M (dim 128, we use the same SIFT1M as online vector) and an industrial dataset DSSMRT40M (dim 64, we have dumped the other 10M online vectors). The specific details of the datasets are as follows: to simulate real-time data insertion, we will replay the SIFT1M dataset for insertion, while the additional 10M data in DSSMRT40M is obtained by dumping newly inserted vectors from an online Kafka source. We have designed the following three sets of experiments to analyze the performance differences between our system and existing systems: comparing the performance differences under different retrieval QPS and real-time insertion QPS on both the standard dataset and the industrial dataset, comparing the performance differences between scenarios with and without rearrangement for high-QPS insertion vectors and comparing the performance differences under different memory block parameters.
The systems we will compare with are as follows:
\begin{itemize}
    \item Faiss \cite{IEEEBILLION-SCALE}: Implements an \texttt{add} interface, which requires copying data back to the CPU for statistical analysis before copying it back to the GPU for actual vector concatenation.
    \item Raft \cite{rapidsai}: Implements an \texttt{extend} interface, allowing reallocation of a new space for the chain on the GPU and swapping.
    \item Rt-cpu: We implement a CPU-based dynamic vector insertion system based on our proposed method, linking pre-allocated memory blocks using linked lists.
    \item RTAMS-GANNS(Proposed Method): The first dynamic vector insertion system for ivfflat and ivfpq on the GPU, supporting multi-stream parallelism with retrieval.
\end{itemize}

To simulate online real-time insertion requests, we have developed a real-time triggering program that triggers a batch of vector insertions at regular intervals. We also set the maximum batch size of Search kernel is 10 since online requests won't be able to wait too long and the cost to batch requests is also expensive. We used a machine equipped with an A10 GPU and 28C-112G AMD Milan processor. The Rt-cpu system only utilizes the CPU part of the processor.
\subsection{Performance on different QPS}
We compared the performance of four systems on the SIFT1M and DSSMRT40M datasets, focusing on latency under high QPS within specific latency constraints, as these are critical for our online systems. We set latency timeouts for both the Insertion and Search Kernels to handle situations when the system cannot process requests in time. Our analysis is confined to the latency observed in the first 10 seconds of operations. To provide a comprehensive view of system performance, we devised a new metric to evaluate the combined latency of the Search and Insertion operations:
\begin{equation}
    \text{latency}_{\text{avg}} = \text{latency}_{\text{avg}}(\text{Search}) + \text{latency}_{\text{avg}}(\text{Insertion})
\end{equation}
We can first conclude that the RTAMS-GANNS (proposed method) not only achieves latency benefits on small datasets (Figure.\ref{fig:compare}abc) but also shows significant improvements on large datasets (Figure.\ref{fig:compare}def). 

Firstly, RTAMS-GANNS can outperform previous systems with up to a 40\% reduction in latency even at low $\text{QPS}_{insertion}$, benefiting from dynamic vector insertion algorithm and multi-stream parallelism. Requests are immediately processed upon arrival without waiting for the previous task to complete, reducing latency by at least 1ms in Figure.\ref{fig:compare}a and d. Secondly, as QPS increases, RTAMS-GANNS has the lowest growth rate in latency, with a maximum reduction of over 80\% (Figure.\ref{fig:compare}ef). Except for a slight increase in Figure.\ref{fig:compare}f due to the higher computational load of large datasets, it shows nearly linear growth under other parameters. Moreover, other systems timeout a lot (reflecting the $\text{latency}_{\text{avg}}$ is over 20ms) when $\text{QPS}_{search}$ is over 5000 and $\text{QPS}_{insertion}$ is over 1000, demonstrating that RTAMS-GANNS's dynamic vector insertion algorithm can efficiently run within the multi-stream parallel framework without causing any delay. 

Among the four systems, the CPU-based system has the highest latency due to its inferior performance compared to the GPU. However, at lower $\text{QPS}_{search}$ on DSSMRT40M, it exhibits a linear increase in latency, which makes it outperform Faiss at $\text{QPS}_{insertion}$ = 2000 in Figure.\ref{fig:compare}d. This observation aligns with our practical experience, where real-time insertions on CPU systems do not significantly affect normal retrieval performance. Additionally, it was found that both the retrieval and insertion performance of Faiss lag behind Raft and RTAMS-GANNS. This is because the add operation in Faiss requires copying the vector list to the CPU and then back to the GPU, which is highly inefficient for large datasets, and the search operation is slower.

\subsection{Affect of Rearrangement}
Secondly, we are particularly concerned about whether the rearrangement mechanism will block subsequent vector insertions and whether it will optimize the performance of online searches. We simulated a scenario where vectors are continuously inserted into the same list and observed the latency fluctuations before and after rearrangement under different rearrangement threshold settings, as well as the degree of performance optimization post-rearrangement. We fixed $\text{QPS}_{search}$ = 5000 and $\text{QPS}_{insertion}$ = 2000 as the request parameters. The experimental results are shown in Table.\ref{tab:rearrage}.

\begin{table}[htbp]
  \caption{Latency Change on Rearrangement Threshold}
  \label{tab:rearrage}
  \begin{tabular}{cccl}
    \toprule
    Threshold&Latency(Before) & Rearrange Cost &Latency(After)\\
    \midrule
    10000 & 4.8ms & 16.1ms & 4.7ms\\
    50000 & 5.1ms & 25.8ms & 5.0ms  \\
    100000 & 7.6ms & 48.1ms & 7.5ms\\
  \bottomrule
\end{tabular}
\end{table}
The threshold values we set in the experiments are particularly large and are not typically triggered to such an extent. We can see that the rearrangement time is within 50ms, which is highly desirable for real-time insertions. This means that once rearrangement is completed, the system immediately updates the vectors that were waiting during this period. Furthermore, after reordering, we observe that the performance of memory reading and jumping is optimized to some extent since the vectors are grouped together. This results in a slight improvement of around 0.1ms for both metrics. While the improvement may seem small, it is still significant.

\subsection{Comparison of Memory Block Parameters}
\begin{figure}[htbp]
\centering
\captionsetup{skip=5pt}
\includegraphics[width=0.75\linewidth]{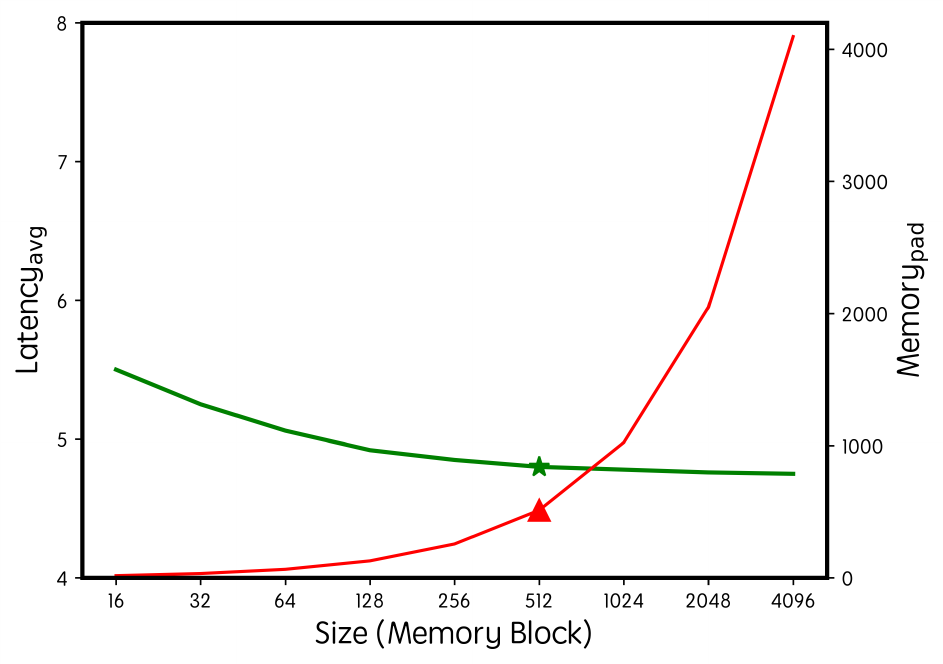}
\caption{Latency/Memory Change on Memory Block Size}
\label{fig:blocksize}
\end{figure}

Finally, we will compare the effects of different memory block sizes in Figure \ref{fig:blocksize}. A larger memory block allows for more vectors to be accommodated, but it also means that a significant amount of memory may be left idle. Conversely, a smaller memory block size may result in frequent memory block allocations, and during searching, there will be more frequent header jumps, which can impact retrieval performance. Similarly, we set $\text{QPS}_{search}$ = 5000 and $\text{QPS}_{insertion}$ = 2000 as constants and solely observe the performance impact of memory block size variations.
We found that as the memory block size increases, the latency decreases. However, the benefits become marginal beyond 1024, and reserving memory blocks larger than 1024 incurs diminishing returns. Additionally, reserving memory blocks of 1024 requires at least 1GB under large datasets if the cluster number of ivf is 4000. Since GPU memory is a precious resource, reserving more than 1GB of pad memory results in significant waste. Moreover, if certain lists are frequently updated with new vectors, this portion of memory will never be utilized. Therefore, we opted for a relatively low-latency option with a pad memory size of 1024, which is acceptable.
\section{Conclusion}
In this paper, we present a Real-Time Adaptive Multi-Stream GPU ANNS System (RTAMS-GANNS). We observed that current GPU ANNS systems primarily focus on offline scenarios, overlooking the high-frequency vector insertion demands in online scenarios. Existing systems face inefficiencies or even global retrieval blocking due to bottlenecks such as the need for new memory allocation and operation within a single stream. Our proposed system overcomes these challenges efficiently by incorporating two innovative components: a memory block based dynamic vector insertion algorithm and a multi-stream execution architecture with stream caching. Experimental results demonstrate that our system effectively handles varying QPS levels across different datasets, reducing latency by up to $40\%-80\%$. Additionally, the experiments show that the dynamic rearrangement mechanism aggregates newly inserted vectors without impacting retrieval, thus reducing latency. We also offer detailed insights into our industry deployment experience, where this system is used in search and recommendation applications, handling over a hundred million user requests daily. Our work demonstrates a stable, reliable, and innovative solution in real-world deployment scenarios.

\balance

\end{document}